\documentclass[prd,preprint,showpacs,showkeys]{revtex4} 

\renewcommand{\vec}[1]{{\bf #1}}       %%%  vectors in bold
\def\beq{\begin{eqnarray}}    %%%  begequation/eqnarray
\def\eeq{\end{eqnarray}}      %%%  endequation/eqnarray

\def\rot{\,\mbox{rot}\,}                %%% rot
\def\al{\alpha}
\def\be{\beta}

\def\ga{\gamma}

\def\pa{\partial}

\def\si{\sigma}

\begin{document}

%%%%%%%%%%%%%%%%%%%%%%%%%%%%%%%%%%%%%%%%%%%%%%%%%%%%%%%%%%%
\title{Exact Foldy-Wouthuysen transformation 
for gravitational waves and magnetic field background}

%%%%%%%%%%%%%%%%%%%%%%%%%%%%%%%%%%%%%%%%%%%%%%%%%%%%%%%%%%%
\author{Bruno Gon\c{c}alves}
\affiliation{Departamento de F\'{\i}sica, ICE,
Universidade Federal de Juiz de Fora,
Juiz de Fora, CEP: 36036-330, MG,  Brazil}

%%%%%%%%%%%%%%%%%%%%%%%%%%%%%%%%%%%%%%%%%%%%%%%%%%%%%%%%%%%
\author{Yuri N. Obukhov}
%\email{yo@thp.uni-koeln.de}
\affiliation{Institute for Theoretical Physics, University of Cologne, 
Z\"ulpicher Stra\ss e 77, 50973 K\"oln, Germany}
\altaffiliation[Also at ]{Department of Theoretical Physics, 
Moscow State University, 117234 Moscow, Russia}

%%%%%%%%%%%%%%%%%%%%%%%%%%%%%%%%%%%%%%%%%%%%%%%%%%%%%%%%%%%
\author{Ilya L. Shapiro}
\affiliation{ Departamento de F\'{\i}sica, ICE,
Universidade Federal de Juiz de Fora,
Juiz de Fora, CEP: 36036-330, MG,  Brazil}

%%%%%%%%%%%%%%%%%%%%%%%%%%%%%%%%%%%%%%%%%%%%%%%%%%%%%%%%%%%
\begin{abstract}
We consider an exact Foldy-Wouthuysen transformation for the Dirac 
spinor field on the combined background of a gravitational wave and 
constant uniform magnetic field. By taking the classical limit of 
the spinor field Hamiltonian we arrive at the equations of motion 
for the non-relativistic spinning particle. Two different kinds of 
the gravitational fields are considered and in both cases the effect 
of the gravitational wave on the spinor field and on the corresponding 
spinning particle may be enforced by the sufficiently strong magnetic 
field. This result can be relevant for the astrophysical applications 
and, in principle, useful for creating the gravitational wave detectors 
based on atomic physics and precise interferometry. 
\end{abstract}
%%%%%%%%%%%%%%%%%%%%%%%%%%%%%%%%%%%%%%%%%%%%%%%%%%%%%%%%%%%%%

\maketitle
%%%%%%%%%%%%%%%%%%%%%%%%%%%%%%%%%%%%%%%%%%%%%%%%%%%%%%%%%%%%%
%%%%%%%%%%%%%%%%%%%%%%%%%%%%%%%%%%%%%%%%%%%%%%%%%%%%%%%%%%%%%
%%%%%%%%%%%%%%%%%%%%%%%%%%%%%%%%%%%%%%%%%%%%%%%%%%%%%%%%%%%%%
\section{Introduction}

The study of the gravitational effects on quantum-mechanical 
systems represents an important issue, since all the physical 
objects, on both macroscopic and microscopic 
scales, are indeed located in a curved spacetime or in 
non-inertial reference frames. This subject is currently of a
special interest (see, e.g., an overview \cite{kiefer}). 
Despite the weakness of the gravitational interaction, its  
effects were actually observed at the quantum-mechanical level. 
In particular, one can mention the famous Colella-Overhauser-Werner 
(COW) \cite{cow} and Bonse-Wroblewski \cite{bonse} experiments, 
in which the quantum-mechanical phase shift due to the 
gravitational and inertial forces was measured, thereby verifying 
the validity of the equivalence principle for the non-relativistic 
neutron waves. 

Among the different configurations of the gravitational field, the 
case of the gravitational waves appears to be especially interesting
\cite{weber}. This is related to the fact that considerable efforts 
are applied to the experimental search of the gravitational waves 
(see, e.g., \cite{gw}). Therefore, it may be important to achieve 
better understanding of the behavior of the quantum fields and 
particles under the action of the gravitational wave. 

In this paper, we study the dynamics of the Dirac particle in a 
plane gravitational wave. The plane-fronted gravitational waves 
represent an important class of exact solutions which generalize 
the basic properties of electromagnetic waves in flat spacetime 
to the case of curved spacetime geometry. The relevant 
investigation of the gravitational waves in general 
relativity has a long and rich history, see, e.g., \cite{exact}. 
On the other hand, the approximate plane gravitational wave 
solution, that arises from the linearized Einstein's field 
equations, is usually used for the analysis of the various 
physical aspects related to the radiation, propagation, and 
detection of the gravitational waves. We will consider both 
cases, the approximate weak wave and the exact nonlinear 
plane-fronted wave. 

The general framework for the Dirac theory in curved spacetime 
was developed in many publications, see the early reference 
\cite{OT}, and the overview and the reference given in a recent 
paper \cite{diraceq}. We will use the notation and conventions 
of the latter work. 

The paper is organized as follows. In section \ref{FWgen} we construct 
the universal form of an exact Foldy-Wouthuysen transformation
for Dirac spinors and apply this scheme to the case of the 
background approximate plane gravitational wave and 
electromagnetic field. Several simpler particular (previously 
well known) cases are considered in Appendix. In section 
\ref{Heom} we use the result of the previous section to consider the  
particle Hamiltonian and corresponding equations of motion 
for the non-relativistic spinning particle interacting to 
the combined background of gravitational wave and electromagnetic 
field. In section \ref{exactwave} we provide the comparison with 
the exact gravitational wave case and in the last section 
\ref{conc} we draw our conclusions.

%%%%%%%%%%%%%%%%%%%%%%%%%%%%%%%%%%%%%%%%%%%%%%%%%%%%%%%%%%%%%
%%%%%%%%%%%%%%%%%%%%%%%%%%%%%%%%%%%%%%%%%%%%%%%%%%%%%%%%%%%%%
%%%%%%%%%%%%%%%%%%%%%%%%%%%%%%%%%%%%%%%%%%%%%%%%%%%%%%%%%%%%%
\section{Linear gravitational perturbations and universal 
form of an exact Foldy-Wouthuysen transformation}\label{FWgen}

Let us start from the relatively simple case of the usual 
(linear) gravitational waves. Using this case as an 
example, we shall also develop the general form of the 
exact Foldy-Wouthuysen transformation which can be applied
to many previously explored cases and also to the more 
complicated case of the exact plane-fronted gravitational wave.

The metric of the weak gravitational wave reads (see, e.g. \cite{MTW})
\beq
g_{ij} = \eta_{ij} + h_{ij}\,,
\label{perturbation}
\eeq
where $\eta_{ij}\,=\,\mbox{diag}(-1,\,1,\,1,\,1)$ is the flat
Minkowski metric and the nonzero components of the gravitational 
perturbation $h_{\mu\nu}$ are (in the Cartesian local coordinates
$x^i = (t,x,y,z)$)
\beq
h_{yy}=-h_{zz}=-2v\,,\qquad h_{yz}=h_{zy}=-2u\,.
\label{wave}
\eeq 
Here $v=v(ct-x)$ and $u= u(ct-x)$ are the two functions which 
describe a wave propagating along the $x$ axis. It is assumed 
that the functions $v=v(ct-x)$ and $u=u(ct-x)$ are small such 
that the linear approximation in these functions is valid. 
As usual, the gravitational wave can have 2 polarization states, 
and each of the functions $v$ and $u$ correspond to one of the 
possible polarizations.

For the formulation of the Dirac theory in curved spacetime, 
we need the tetrad fields. The coframe 1-form reads:
\begin{eqnarray}
\vartheta^0 &=& cdt,\qquad \vartheta^1 = dx,\\
\vartheta^2 &=& (1 + v)dy + udz,\\
\vartheta^3 &=& (1 - v)dz + udy.
\end{eqnarray}
The corresponding inverse vector frame is described by
\begin{eqnarray}
e_0 &=& {\frac 1 c}\partial_t,\qquad e_1 = \partial_x,\\
e_2 &=& (1 - v)\partial_y - u\partial_z,\\
e_3 &=& (1 + v)\partial_z - u\partial_y.
\end{eqnarray}
It is straightforward to construct the Riemannian connection (the 
Christoffel symbols). The nonzero components of the local connection 
1-forms read:
\begin{eqnarray}
\Gamma_2{}^0 
&=& \Gamma_0{}^2 = v'\,\vartheta^2 + u'\,\vartheta^3,
\label{g1}
\\
\Gamma_3{}^0 
&=& \Gamma_0{}^3 = u'\,\vartheta^2 - v'\,\vartheta^3,
\\
\Gamma_2{}^1 
&=& 
- \Gamma_1{}^2 = v'\,\vartheta^2 + u'\,\vartheta^3,
\\
\Gamma_3{}^1 
&=& 
- \Gamma_1{}^3 = u'\,\vartheta^2 - v'\,\vartheta^3.
\label{g4}
\end{eqnarray}
The primes denote the derivatives w.r.t. the argument of the 
functions: $v' = dv(z)/dz, u' = du(z)/dz$.

One can check that the Cartan structure equation is fulfilled.
Indeed, we have $d\vartheta^\alpha + \Gamma_\beta{}^\alpha\wedge
\vartheta^\beta =0$, which demonstrates that this is the torsion-free 
Riemannian connection. 

Let us construct the Dirac operator on the background of the 
metric described above. The spinor covariant derivative is
\begin{equation}
D_\alpha = e_\alpha\rfloor D,\qquad 
D := d + {\frac i 4}\,\widehat{\sigma}^{\alpha\beta}
\,\Gamma_{\alpha\beta}.\label{Dspinor}
\end{equation}
We have 
$\widehat{\sigma}^{\alpha\beta}= i\gamma^{[\alpha}\gamma^{\beta]}$,
and consequently, 
\begin{equation}
{\frac i 4}\,\widehat{\sigma}^{\alpha\beta}
\,\Gamma_{\alpha\beta} 
= -\,{\frac 1 2}\left(\gamma^2\gamma^0\,\Gamma_{20}
+ \gamma^3\gamma^0\,\Gamma_{30} 
+ \gamma^1\gamma^2\,\Gamma_{12}
+ \gamma^3\gamma^1\,\Gamma_{31}\right). 
\end{equation}
We shall use the standard \cite{Bjorken-Drell} representation 
for the gamma matrices:
\begin{equation}
\gamma^0 = \beta 
= \left(\begin{array}{cc}{\rm I} & 0\\ 0 & -{\rm I}
\end{array}\right),\qquad\gamma^a 
= \left(\begin{array}{cc} 0 & \sigma^a\\
-\sigma^a & 0 \end{array}\right),\quad a = 1,2,3,
\label{gammas}
\end{equation}
where the Pauli matrices are:
\begin{equation}
\sigma^1 = \left(\begin{array}{cc}0&1\\ 1&0\end{array}\right),
\qquad
\sigma^2 = \left(\begin{array}{cr}0&-i\\ i&0\end{array}\right),
\qquad
\sigma^3 = \left(\begin{array}{cr}1&0\\ 0&-1\end{array}\right).
\label{pauli}
\end{equation}
The alpha matrices and the spin matrix are defined, as usual, 
by the relations
\begin{equation}
\overrightarrow{\alpha} 
= \gamma^0\overrightarrow{\gamma} = \left(\begin{array}{cc} 0 &
\overrightarrow{\sigma}\\ \overrightarrow{\sigma}&0\end{array}\right),
\qquad
\overrightarrow{\Sigma} = \left(\begin{array}{cc}\overrightarrow{\sigma} & 0
\\ 0& \overrightarrow{\sigma}\end{array}\right).
\end{equation}
Using these definitions, and substituting 
(\ref{g1})-(\ref{g4}), we find
\begin{equation}
{\frac i 4}\,\widehat{\sigma}^{\alpha\beta}
\,\Gamma_{\alpha\beta} 
= {\frac 1 2}\left(\alpha^2 + i\,\Sigma^3\right)
\left(v'\,\vartheta^2 + u'\,\vartheta^3\right) 
+ {\frac 1 2}\left(
\alpha^3 - i\,\Sigma^2\right)\left(u'\,\vartheta^2 
- v'\,\vartheta^3\right).
\end{equation}

The Dirac operator is constructed as $\gamma^\mu D_\mu$, and
correspondingly, we obtain
\begin{eqnarray}
\gamma^\mu e_\mu\rfloor 
\left({\frac i 4}\,\widehat{\sigma}^{\alpha\beta}
\,\Gamma_{\alpha\beta}\right) 
&=& 
{\frac {v'} 2}
\left[\left(\gamma^2\alpha^2 - \gamma^3\alpha^3\right) 
+ i\left(\gamma^2\Sigma^3 
+ \gamma^3\Sigma^2\right)\right]\nonumber\\ 
&& 
+ \,{\frac {u'} 2}
\left[\left(\gamma^2\alpha^3 + \gamma^3\alpha^2\right) 
+ i\left(\gamma^3\Sigma^3 - \gamma^2\Sigma^2\right)\right].
\label{affine}
\end{eqnarray}
A direct computation shows that 
\begin{equation}
\gamma^2\alpha^2 - \gamma^3\alpha^3 = 
\gamma^2\alpha^3 + \gamma^3\alpha^2 = 
\gamma^2\Sigma^3 + \gamma^3\Sigma^2 = 
\gamma^3\Sigma^3 - \gamma^2\Sigma^2 = 0.
\end{equation}
Hence, all the terms on the r.h.s. of the equation 
(\ref{affine}) do vanish .
As a result, the contribution of the spinor connection drops 
out completely and the Dirac operator is reduced just to the 
partial derivative terms
\begin{eqnarray}
\gamma^\mu D_\mu 
= \gamma^\mu e_\mu\rfloor d &=& {\frac 1 c}\,\gamma^0
\,\partial_t + \gamma^1\,\partial_x 
+ \left[(1 - v)\,\gamma^2 - u\,\gamma^3
\right]\partial_y \nonumber\\
&& 
+\,\left[(1 + v)\,\gamma^3 
- u\,\gamma^2\right]\partial_z\nonumber
\\
&=& \beta\,\Big\{{\frac 1 c}\,\partial_t 
+ \alpha^1\,\partial_x 
+ \left[(1 - v)\,\alpha^2 
- u\,\alpha^3\right]\partial_y 
\nonumber
\\
&& +\,\left[(1 + v)\,\alpha^3 
- u\,\alpha^2\right]\partial_z\Big\}.
\end{eqnarray}

Now we are in the position to consider the covariant Dirac equation
\begin{equation}
\left(i\hbar \gamma^\mu D_\mu - mc\right)\psi = 0\,.
\label{Dirac0}
\end{equation}
After we recast it into the familiar Schr\"odinger form,
this equation reduces to
\begin{equation}
\label{Dirac1}
i\hbar{\frac {\partial\psi} {\partial t}} = \widehat{\cal H}\,\psi
\end{equation}
with the Hamilton operator of the form
\begin{equation}
\widehat{\cal H} = mc^2\beta + c\alpha^1\,p^1 + 
\left[(1 - v)\,c\alpha^2 - u\,c\alpha^3\right] p^2
+ \left[(1 + v)\,c\alpha^3 - u\,c\alpha^2\right] p^3\,,
\end{equation}
where $\,p^1,\,p^2,\,p^3\,$ are components of the momentum vector
$\,\overrightarrow{p}$. 

The crucial observation is that this form of the Hamiltonian 
falls into the class of models which admit the anticommuting 
involution operator, cf. \cite{diraceq,EK,nikitin}, 
\begin{equation}
J = i\gamma_5\beta.
\label{invol}
\end{equation}
The latter is Hermitian, $J^\dagger = J$, and unitary, 
$JJ^\dagger = J^2 = 1$, 
and it anticommutes both with the Hamiltonian  and with the 
$\beta$ matrix:
\begin{equation}
J\widehat{\cal H} + \widehat{\cal H}J = 0,\qquad 
J\beta + \beta J = 0.\label{involcom}
\end{equation}
Consequently, the exact FW transformation can be constructed 
in this case.

Moreover, if one introduces the (in the simplest case, constant 
and uniform) magnetic field, this important feature is not 
destroyed, since the only thing which is technically needed 
is to replace the operators of momentum $\overrightarrow{p}$ 
by the expressions $\overrightarrow{p} - {\frac ec}\overrightarrow{A}$. 
After performing this operation we arrive at the following
Hamiltonian:
\beq
H &=& \be mc^2 - i\hbar c\Big[\alpha^1(\pa_x+\frac {e} {i\hbar c} A_x)
\label{H}
\\
&+& (\alpha^2 - v\alpha^2-u\alpha^3)(\partial_y
+ \frac {e} {i\hbar c} A_y)
+ (\alpha^3 -u\alpha^2+v\alpha^3)(\partial_z
+\frac {e}{i\hbar c} A_z)\Big]\,.
\nonumber
\eeq
For  a while we do not impose restrictions for the vector potential. 
Let us rewrite the expression (\ref{H}) using new notations, 
which prove useful in what follows \cite{notations}
\beq
H\,=\,\be mc^2 +\,\alpha^b K^a_b\partial_a+\,\alpha^ag_a\,,
\label{not}
\eeq
where
\beq
K^a_b=-i\hbar c \pmatrix{1   &     0      &        0  \cr       
0   & 1-v-u\alpha^2\alpha^3  &  0                     \cr        
0   &   0      & 1+v-u\alpha^3\alpha^2                \cr
} \,,
\eeq
\beq
g_a= \,-e\,(A_1\,,\, A_2-vA_2-uA_3\,,\, A_3-uA_2+vA_3)\,,
\eeq
According to the standard prescription \cite{EK} 
(the structure of exact FW for the scalar fields with non-minimal 
coupling to gravity has been discussed recently in \cite{accioly})
the fist step in deriving the exact FW is to calculate the square 
of the Hamiltonian $H^2$. In order to accomplish this, 
we define, additionally, conjugated quantities
\beq
\overline{K^a_b}=-i\hbar c\pmatrix{1 & 0  &  0  \cr
0    &  1-v+u\alpha^2\alpha^3  &   0            \cr 
0    &  0    & 1+v+u\alpha^3\alpha^2            \cr}\,,
\eeq
\beq
\overline{g_a}
\,= \,(\alpha^1g_1\,,\, \alpha^2g_2\,,\, \alpha^3g_3\,)\,,
\eeq
and 
\beq
\overline{\epsilon^{abc}}= \,(\alpha_1\epsilon^{a1c}\,,\, 
\alpha_2\epsilon^{a2c}\,,\, \alpha_3\epsilon^{a3c})\,.
\eeq

Using these notations, a direct calculation gives, after 
some algebra, the following square of the Hamiltonian \ $H^2$:
\beq
H^2 &=& m^2c^4\,+\,\overline{K^{ac}}K^b_c\pa_a\pa_b\,
+\,\overline{K^a_b}g^b\pa_a+g^bK^a_b\pa_a\,+\,g^2\, \nonumber \\
&+& \overline{K^{ac}}(\pa_a K^b_c)\pa_b
\,+\,i\Sigma_d\epsilon^{bcd}\overline{K^a_b}(\pa_a K^e_c)\pa_e
\,+\,i\Sigma_d \overline{\epsilon^{bcd}}K^a_b\overline{g_c}\pa_a
\nonumber \\
&-& 
i\epsilon^{bcd}\Sigma_d K^a_b g_c\pa_a
\,+\,i \Sigma_d \overline{\epsilon^{bcd}}K^a_b(\pa_a \overline{g_c})
\,+\,\overline{K^a_b}(\pa_a g^b) \,.
\label{resultado}
\eeq
Here the partial derivatives inside the parenthesis act only 
on the content of these parenthesis.
Let us notice that the  components of the matrix $K^a_b$ depend 
exclusively 
on the metric while the  components of the vector $g_a$ depend on both 
the metric and the electromagnetic potential $A_a$.
The expression (\ref{resultado}) is rather general and may be 
used not only for the linear gravitational waves but also in 
various other particular cases. In order to illustrate this 
fact, we consider several known cases in the Appendix.

%%%%%%%%%%%%%%%%%%%%%%%%%%%%%%%%%%%%%%%%%%%%%%%%%%%%%%%%%%%
\section{Particle Hamiltonian and equations of motion}\label{Heom}

Starting from the equation (\ref{resultado}), for the sake of 
simplicity, we consider only one polarization of the gravitational
wave. Namely, we choose $u=0$. Then, we obtain
\beq
K^a_b=\overline{K^a_b}=-i\hbar c(\delta^a_b+T^a_bv)
\qquad,\qquad g_b = - e(\delta^a_b+T^a_bv)A_a\,,
\eeq
where
\beq
T^a_b= \pmatrix{0 &0 &0 \cr 0 &-1 &0 \cr 0 &0 &1 \cr}\,,\label{T}
\eeq
and there is no need anymore to use the notations 
\ $\overline{g_b}$ \ and \ $\overline{\epsilon^{abc}}$. 
Thus we arrive at the Hamiltonian
\beq
H &\simeq& \frac{1}{2mc^2}\,\be\,
\big(\delta^{ab} + 2T^{ab} v\big)\,
\Big[(cp_a-eA_a)(cp_b-eA_b)-
e\hbar c\,\varepsilon_{cad}\Sigma^d\pa^c(A_b)\Big]\nonumber\\
&+& \frac{\hbar}{2mc}\,\be\,\varepsilon^{abc}\pa_a(v)\Sigma_c 
T^d_b(cp_d 
- e A_d) + \be mc^2\,.\label{htransf}
\eeq

%%%%%%%%%%%%%%%%%%%%%%%%%%%%%%%%%%%%%%%%%%%%
\subsection{Nonrelativistic limit}

The next step is to present the Dirac fermion $\psi$ in the 
form
\beq
\psi  = 
\left( \begin{array}{c}\varphi \\ \chi \end{array}\right)
e^{-imc^2t/\hbar}\;\;,\label{expo}
\eeq
and use the equation 
\beq
i\hbar\pa_t\psi=H\psi
\label{S}
\eeq
to derive the Hamiltonian for the 2-spinor $\varphi$. Inserting 
(\ref{expo}) into (\ref{S}), we obtain the two-component equation
\beq
i\hbar\frac{\pa}{\pa t}\left( \begin{array}{c} \varphi \\ 
\chi\end{array}
\right)\, = \,\left(- mc^2+H\right)\,\,\left( \begin{array}{c} \varphi 
\\ \chi \end{array}\right)\;.\label{eqexpo}
\eeq
Using the fact that Hamiltonian is an even function,
we obtain, in the $\varphi$ sector, the Hamiltonian
\beq
H &=& \frac{1}{2mc^2}\,\big(\delta^{ab} + 2T^{ab} v\big)
\,\,\Big[(cp_a - eA_a)(cp_b - e A_b) - e\hbar c \varepsilon_{cad}
\sigma^d\pa^cA_b\Big]\nonumber\\
&+& \frac{\hbar}{2mc}\varepsilon^{abc}\pa_a(v)\sigma_c 
T^d_b(cp_d-eA_d)\;.
\label{hfi}
\eeq

The r.h.s. of the last equation (\ref{hfi}) is 
proportional to \ $1/m$, that is the same level of the 
nonrelativistic approximation which one meets in the 
case of Pauli equation. Therefore one can expect that 
the same equation can be obtained starting from the 
original eq. (\ref{not}). This calculation represents 
an efficient check of our results, so it is worthwhile 
to perform it now. Let us apply the standard procedure 
for deriving the Pauli equation (see, e.g. \cite{LL-4}). 
Using the representation (\ref{eqexpo}) in (\ref{not})
and (\ref{S}) we apply the low-energy regime , that is 
assume that the term $mc^2$ is dominant
$(|mc^2\chi|\gg|i\hbar\pa_t\chi|)$. In this way we arrive 
at the equation
\beq
i\hbar\frac{\pa}{\pa t}\varphi=
\frac{1}{2mc^2}(\sigma^bK^a_b\pa_a+\sigma^ag_a)
(\sigma^dK^c_d\pa_c+\sigma^cg_c)\varphi\,.
\eeq
After some transformations, the r.h.s. of the last 
equation coincides with the r.h.s. of our eq. 
(\ref{hfi}).

It is easy to check that, if the gravitational wave is 
absent \ $v\equiv0$, one obtains 
\beq
i\hbar\frac{\pa}{\pa t}\varphi=
\frac{1}{2mc^2}
[\overrightarrow{\sigma}
(c\overrightarrow{p}-e\overrightarrow{A})\cdot
\overrightarrow{\sigma}
(c\overrightarrow{p}-e\overrightarrow{A})]\varphi\;\;,
\eeq
and finally, after a some algebra, a usual Pauli 
equation
\beq
i\hbar\frac{\pa }{\pa t}\varphi =
[\frac{1}{2m}
(\overrightarrow{p}-\frac{e}{c}\overrightarrow{A})^{2}
-\frac{e\hbar}{2mc}
\overrightarrow{\sigma}\cdot\overrightarrow{B}] \varphi\;.
\label{paulih}
\eeq

%%%%%%%%%%%%%%%%%%%%%%%%%%%%%%%%%%%%%%%%%%%%
\subsection{Nonrelativistic spinor particle on the background
of the gravitational wave and electromagnetic field}

One can use the the result (\ref{htransf}) for deriving the 
equation for the nonrelativistic spinor particle. Let us 
follow \cite{BBS,torsi}, where the similar calculation has
been performed for the nonrelativistic equation on the 
background of torsion field. The same problem has been 
treated in \cite{rytor} starting from the perturbative 
FW transformation. 

The Hamiltonian operador $\hat{H}$ corresponding to the
energy (\ref{htransf}) is constructed in terms
of the operators \ $\hat{x}_a$, \ $\hat{p}_a$ \ and 
\ $\hat{\sigma}_a$. The equations of motion have the 
form 
\beq
i\hbar\frac{d\hat{x}_a}{dt}=\big[\hat{x}_a,H\big]
\,, \qquad
i\hbar\frac{d\hat{p}_a}{dt}=\big[\hat{p}_a,H\big]
\,, \qquad
i\hbar\frac{d\hat{\sigma}_a}{dt}=\big[\hat{\sigma}_a,H\big]\;.
\label{eqm}
\eeq
In order to achieve the nonrelativistic limit one has 
to calculate the commutators of the operators in (\ref{eqm})
and take the limit $\hbar\rightarrow0$.
Disrearding the $\,{\cal O}(\hbar)\,$ terms, we do not 
need to care about the ordering of the operators, and then
finally we arrive at the semiclassical equations of motion 
for the spinning particle on the background of the gravitational
field and constant magnetic field. 
The result can be presented using the notation 
$A^{\prime}_a = T_{ab}A^b$. Then the equations become
\beq
\frac{dx_a}{dt} &=& \frac{1}{m}(\delta_{ab} 
+ 2\,T_{ab}\,v)\left(p^b-\frac{e}{c}A^b\right)\;,\label{acel}\\
\nonumber\\
\frac{dp_a}{dt}&=& - 
\frac{1}{m}\,T^{bc}\,\pa_av\,\left(p_b-\frac{e}{c}A_b
\right)\left(p_c-\frac{e}{c}\,A_c\right)\,+\,(\delta_{bc} + 2T_{bc}v)
\,\left(p^b-\frac{e}{c}A^b\right)\,\frac{e}{mc}\,\pa_a 
A^c\,,\label{dP}\\ 
\nonumber \\
\frac{d\sigma_a}{dt} &=& \frac{e}{mc}\,\varepsilon_{abc}\sigma^c
\left[\overrightarrow{B} + 2v \rot(\overrightarrow{A}^\prime)\right]^b
\,-\,\frac{1}{mc}\,(cp^d-eA^d)\,\sigma^e\,\left[T_{ed}(\pa_a v)
-T_{ad}(\pa_e v)\right]\,.\label{eqmexp}
\eeq   
Let us notice that the first equation here demonstrates 
deviation from the usual relation between the Lagrangian 
velocity and the canonically conjugated momenta. The 
difference is due to the gravitational wave. Taken together,
the first two equations define the coordinate 
dependence of the particle. 
Using these two equations, one can construct the 
analog of Lorentz force in the presence of gravitational wave
\beq
m\,\ddot{x}_a &=& \frac{e}{c}\,(\delta_{ab}+2T_{ab}v)\Big[
\dot{\overrightarrow{x}}\times  \overrightarrow{B}]^b 
- \frac{e}{c}(\delta_{ab}+2T_{ab}v)\frac{\pa A^b}{\pa t}\nonumber\\
&+& 2mT_{ab}\frac{dv}{dt}\dot{x}^b
- mT_{bc}\,\dot{x}^b\,\dot{x}^c(\pa_a v)\,.\label{xdd}
\eeq
The last equation  in (\ref{eqmexp}) in describing the spin 
dynamics of the particle. It is remarkable that this 
equation depends on the velocity. Let us notice that the 
same property holds for the spinning particle in the 
external torsion field for both nonrelativistic \cite{BBS}
and relativistic \cite{gegi} cases.    

Perhaps the spin dynamics of the particle is the most 
interesting result here. In order to understand this point, 
let us make the following observation. It is easy to notice 
that the last equation  in (\ref{eqmexp}) describes spin 
precession even for the case when the magnetic field is 
absent. The effect is due to the presence of the matrix 
$T^a_b$ defined in eq.(\ref{T}) and the $\rot {\vec A}^\prime$. 
The last vector is distinct from the magnetic field and maybe 
nonzero when ${\vec B}=\rot {\vec A}=0$. Therefore, there is a 
possibility to observe the spin precession without magnetic 
field and this can, in principle, become a new basis for the 
gravitational waves detector of the new type. We expect 
to consider this issue in more details in the near future. 
A detailed analysis of the definitions of the momentum and
spin dynamical operators (along the lines recently done for
the static gravitational field \cite{ST}) will be required. 

%%%%%%%%%%%%%%%%%%%%%%%%%%%%%%%%%%%%%%%%%%%%%%%%%%%%%%%%%%%%%
\subsection{Vanishing electromagnetic field}

In the absence of the electromagnetic potential, $A_a = 0$, the 
equations of motion can be intergrated. With $x^a = (x, y, z)$, 
the second and the third equations in (\ref{xdd}) read 
\begin{eqnarray}
{\frac d {dt}}\,\dot{y} &=& -\,2\dot{y}\,{\frac {dv} {dt}},\label{yd}\\
{\frac d {dt}}\,\dot{z} &=& \,2\dot{z}\,{\frac {dv} {dt}}.\label{zd}
\end{eqnarray}
These equations are easily integrated, yielding the components of the
velocities as the functions of $v$:
\beq
\dot{y} = Y\,e^{-2v},\qquad \dot{z} = Z\,e^{2v}\,,
\label{yzd}
\eeq
with the integration constants $Y$ and $Z$. The integration of the 
equation
for $x$ is slightly more nontrivial. This equation reads
\begin{equation}
\label{eqx}
\ddot{x} = (\dot{y}^2 - \dot{z}^2)\,\partial_x v = f(v)\,\partial_x v,
\end{equation}
with $f(v) = Y^2e^{-4v} - Z^2e^{4v}$. Recalling that $v = v(\xi)$ with
$\xi = ct - x$, we notice that $\partial_x v = -\,dv/d\xi$, and 
furthermore
$\ddot{\xi} = - \ddot{x}$. Accordingly, we can recast (\ref{eqx}) into 
the
equation for $\xi$:
\begin{equation}
2\ddot{\xi} = -\,{\frac {dU}{d\xi}},\qquad U = {\frac 
12}\left(Y^2e^{-4v} 
+ Z^2e^{4v}\right). 
\end{equation}
Multiplying with $\dot{\xi}$, we find the first integral
\begin{equation}
\dot{\xi}^2 + U(\xi) = I,
\end{equation}
and the solution $\xi = \xi(t)$ is then obtained in quadratures 
\beq
\int {\frac {d\xi}{\sqrt{I - U(\xi)}}} = t - t_0\,. 
\label{x}
\eeq
The explicit form of the solution $\xi = \xi(t)$ of course depends on 
the
integration constant $I$ and on the explicit form of the wave function 
$v = v(\xi)$ that will determine the ``potential" $U(\xi)$. For the 
harmonic
wave, for example, $v = v_0\,\cos\xi$, and then $U(\xi) = \tilde{Y}^2
e^{-4\cos\xi} +  \tilde{Z}^2e^{4\cos\xi}$ (with some new constants 
$\tilde{Y}$ and $\tilde{Z}$). After finding $\xi(t)$, we can use it for
the final integration of the first order equations (\ref{yzd}). Thus, 
we
finally obtain the coordinates of the particle as functions of time 
$y(t)$, $z(t)$ and $x(t) = ct - \xi(t)$. 

In order to complete the analysis of the particle dynamics, we have to
solve the equation for the spin. In the absence of the electromagnetic
field, it reads:
\begin{equation}
\dot{\sigma}_a = {\frac {\sqrt{p_y^2 + p_z^2}\,(\partial_x v)}{m}}
\,M_{ab}\,\sigma_b,
\end{equation}
with the matrix 
\beq
M_{ab} = \left(\begin{array}{ccc}0 & \pi_y & - \pi_z \\ 
-\pi_y & 0 & 0 \\ \pi_z & 0 & 0\end{array}\right).
\eeq
Here $\pi_y = p_y/\sqrt{p_y^2 + p_z^2}$, $\pi_z = p_z/\sqrt{p_y^2 + 
p_z^2}$.
Note that $\dot{p_y} = \dot{p_z} = 0$, see (\ref{dP}). The integration 
is 
then straightforward, yielding the final result for the dynamics of the 
spin
\beq
\overrightarrow{\si}\,=\,\exp \left[s(t) M\right]\cdot
\overrightarrow{\si}_0\,,\qquad \mbox{where} \qquad  s(t) = {\frac 
{\sqrt{p_y^2 + p_z^2}}m}\int\limits_{t_0}^t dt\,\partial_x 
(v)\,.\label{spin}
\eeq
Note that since for the cubic term we have $M^3 = - \,M$, the matrix 
exponential actually contains only the terms $M$ and $M^2$, and it can
be written explicitly as 
$$
%%  \exp\left[s(t) M\right] 
e^{s(t) M} = {\bf 1} + M\sin s 
+ M^2\left(1 - \cos s\right)\,.
$$ 

The solutions (\ref{yzd}), (\ref{x}), (\ref{spin}) show that the 
nonrelativistic spinning particle has very peculiar behaviour in 
the field of the weak gravitational wave.

%%%%%%%%%%%%%%%%%%%%%%%%%%%%%%%%%%%%%%%%%%%%
%%%%%%%%%%%%%%%%%%%%%%%%%%%%%%%%%%%%%%%%%%%%
%%%%%%%%%%%%%%%%%%%%%%%%%%%%%%%%%%%%%%%%%%%%
\section{Comparison with exact gravitational wave}\label{exactwave}

The exact plane-fronted gravitational wave, in the simplest case
is described in the Cartesian local coordinates $x^i = (t,x,y,z)$ by 
the line element $ds^2 = g_{ij}dx^idx^j$ with  
the metric (see \cite{peres,exact,pp})
\beq
g_{ij} = \eta_{ij} + h_{ij}\,,
\label{perturbation2}
\eeq
where the nonzero components are
\beq
h_{tt}=h_{xx} = - U,\qquad h_{tx}=h_{xt}= U\label{wave2}
\eeq 
expressed in terms of a function $U(\xi,y,z)$. It can depend 
arbitrarily
on $\xi = ct - x$, and is a harmonic function in the two last 
variables,
i.e. $\Delta_{(2)}U = (\partial^2_{yy} + \partial^2_{zz})U = 0$. 
The coframe reads
\begin{eqnarray}
\vartheta^0&=&\left(1 + {\frac U2}\right)cdt -{\frac 
U2}\,dx,\label{tetrada2}\\
\vartheta^1&=&{\frac U2}\,cdt + \left(1 - {\frac U2}\right)dx,\\
\vartheta^2&=& dy,\qquad \vartheta^3  = dz.
\end{eqnarray}
The inverse frame is easily found:
\begin{eqnarray}
e_0&=&\left(1 - {\frac U2}\right){\frac 1c}\partial_t 
-{\frac U2}\,\partial_x,\\
e_1&=&{\frac U{2c}}\,\partial_t 
+ \left(1 + {\frac U2}\right)\partial_x,\\
e_2&=& \partial_y,\qquad e_3  = \partial_z.
\end{eqnarray}
{}From this we can verify straightforwardly that
\begin{equation}
{\frac i 4}\,\widehat{\sigma}^{\alpha\beta}\,\Gamma_{\alpha\beta} 
= \left(\gamma^0 - \gamma^1\right){\frac 14}\left(\gamma^2\partial_yU 
+ \gamma^3\partial_zU\right)\left(\vartheta^1 - \vartheta^0\right).
\end{equation}
Accordingly, we find that
\begin{equation}
\gamma^\mu e_\mu\rfloor \left({\frac i 
4}\,\widehat{\sigma}^{\alpha\beta}
\,\Gamma_{\alpha\beta}\right) = -\left(\gamma^0 - \gamma^1\right)^2
{\frac 14}\left(\gamma^2\partial_yU + \gamma^3\partial_zU\right) = 0,
\end{equation}
since $\left(\gamma^0 - \gamma^1\right)^2 \equiv 0$. Thus, just like in
the case of an approximate wave, the spinor connection term drops out
completely from the Dirac equation.

Consequently, the Dirac operator has the form
\beq
\ga^\alpha D_\alpha = \gamma^\alpha e_\alpha\rfloor d = 
{\frac 1c}\Big[\ga^0+\frac{U}{2}(\ga^1-\ga^0)\Big]\pa_t +
\Big[\ga^1+\frac{U}{2}(\ga^1-\ga^0)\Big]\pa_x +
\ga^2\pa_y + \ga^3\pa_z\,.
\eeq 
It proves very useful, before we go to the exact FW transformation,
we make a Lorentz transformation of the coframe $\vartheta^\alpha 
\rightarrow\vartheta'^\alpha = \Lambda^\alpha{}_\beta\vartheta^\beta$, 
using the matrix (written in the $2\times 2$ block form with {\bf 0} 
and
{\bf 1} as the $2\times 2$ zero and unit matrices, respectively)
\beq
\Lambda^\alpha{}_\beta \,=\, \pmatrix{L  &{\bf 0}\cr
{\bf 0}   &{\bf 1} \cr}\,,\qquad\mbox{where}\qquad
L = \frac{1}{\sqrt{1-U}}\pmatrix{1-U/2  &   U/2   \cr
U/2    &  1-U/2  \cr}\,.\label{LO}
\eeq
Under this transformation, the tetrad frame is changed to $e_\alpha
\rightarrow e'_\alpha = (\Lambda^{-1})^\beta{}_\alpha e_\beta$, whereas
the local Lorentz connection transforms to $\Gamma'_\alpha{}^\beta 
= \Lambda^\beta{}_\nu\Gamma_\mu{}^\nu({\Lambda^{-1}})^\mu{}_\alpha
+ \Lambda^\beta{}_\gamma d(\Lambda^{-1})^\gamma{}_\alpha$. One can then 
verify that the transformed spinor connection term do contribute 
to the Dirac operator
\beq
\gamma^\mu e'_\mu\rfloor \left({\frac i 
4}\,\widehat{\sigma}^{\alpha\beta}
\,\Gamma'_{\alpha\beta}\right) = \frac{U'}{4(1-U)^{\frac{3}{2}}}
(\gamma^0 - \gamma^1) + 
\frac{1}{4(1-U)}\,\gamma^0\left(\gamma^2\gamma^1
\pa_yU + \gamma^3\gamma^1\pa_z U\right)\,,
\eeq
where $U' = dU/d\xi$. Another contribution comes from the ordinary 
derivative term
\beq
\gamma^\alpha e'_\alpha\rfloor d = \gamma^0\left({\frac {\sqrt{1-U}} 
c}\pa_t 
- \frac{U}{\sqrt{1-U}}\,\pa_x\right) + 
\gamma^1\,\frac{1}{\sqrt{1-U}}\pa_x 
+ \gamma^2\pa_y +\gamma^3\pa_z\,.
\eeq
Collecting all together, we can finally write the Dirac equation 
(\ref{Dirac0}) in the Schr$\ddot{\rm o}$dinger form (\ref{Dirac1}). 
As a last step we perform the rescaling transformation \cite{OT} of 
the wave function 
$$
\Psi^\prime=(1-U)^{\frac{1}{4}}\Psi\,.
$$ 
In this way we arrive at the final form of the Hamiltonian that is 
explicitly Hermitian:
\beq
H^\prime &=& \beta mc^2V + cp_x - \frac{c}{2}(1-\alpha^1)\left(
V^2p_x + p_x V^2\right)\nonumber \\
&& +\,{\frac{c}{2}}\left[\alpha^2(V p_y + p_y V) + \alpha^3(Vp_z + 
p_zV)
\right]\nonumber\\
&& -\,{\frac{i\hbar c}{2}}\,\alpha^1\left[\alpha^2\pa_y(V) + 
\alpha^3\pa_z(V)\right]\,.\label{Hew}
\eeq
Here $V=1/\sqrt{1-U}$. 

Let us notice that switching the gravitational wave off with $U=0$
(hence $V = 1$), we recover the Hamiltonian of the free 
particle. The Hamiltonian (\ref{Hew}) is not anticommuting with the 
matrix $\,J=i\gamma^5\beta$. As a result, the condition for performing 
the exact FW transformation ($\{J\,,\,H'\}=0$) is not satisfied in 
this case.

%%%%%%%%%%%%%%%%%%%%%%%%%%%%%%%%%%%%%%%%%%%%%%%%%%%%%%%%%%%
\section{Discussion and conclusion}\label{conc}

In this paper, we have derived an exact Foldy-Wouthuysen
transformation for the Dirac spinor field on the combined 
background of a gravitational wave and constant uniform magnetic 
field. The motivation for the presence of the magnetic field
is to check the possibility of the amplification of the 
influence of the gravitational wave on a Dirac particle. 
According to our calculations (\ref{yzd}), (\ref{dP}) and 
(\ref{eqmexp}) such an effect is possible. This result can be
relevant 
for the astrophysical applications and, in principle, could be 
useful for improving the gravitational wave detectors based on 
atomic physics and precise interferometry (see, e.g., 
recent discussion in \cite{vit}). The actual dynamics
of the Dirac particle in a plane gravitational wave will be
studied separately. 

\bigskip
{\bf Acknowledgments}. The work of B.G. and I.Sh. has been 
supported by the PRONEX projects and research grants from 
FAPEMIG (MG, Brazil) and CNPq (Brazil) and also (I.Sh.) by 
the PRONEX project from UFES (ES, Brazil) and ICTP (Italy).
For Y.N.O. this work was partially supported by FAPESP 
(S\~{a}o Paulo, Brazil) and by DFG (Bonn). 
%%%%%%%%%%%%%%%%%%%%%%%%%%%%%%%%%%%%%%%%%%%%%%%%%%%%%%%%%%%
%%%%%%%%%%%%%%%%%%%%%%%%%%%%%%%%%%%%%%%%%%%%%%%%%%%%%%%%%%%

\appendix

%%%%%%%%%%%%%%%%%%%%%%%%%%%%%%%%%%%%%%%%%%%%%%%%%%%%%%%%%%%
%%%%%%%%%%%%%%%%%%%%%%%%%%%%%%%%%%%%%%%%%%%%%%%%%%%%%%%%%%%
\section{Free particle}

Let us consider several simple known particular cases when the 
general formula (\ref{resultado}) can be applied.

For the case of a free particle, we have \ 
$K^a_b=-i\,\hbar\,c\,\delta^a_b$ 
and $g_a=0$, because $u=v=0$ and $A_b=0$.
Then
\beq
H=\beta \sqrt{m^2c^4+c^2p^2}\,,
\eeq
that is a well-known known textbooks result.

%%%%%%%%%%%%%%%%%%%%%%%%%%%%%%%%%%%%%%%%%%%%%%%%%%%%%%%%%%%
%%%%%%%%%%%%%%%%%%%%%%%%%%%%%%%%%%%%%%%%%%%%%%%%%%%%%%%%%%%
\section{Particle in a magnetic field}

In this case \ $K^a_b=-i\,\hbar\,c\,\delta^a_b$
and $g_a=-eA_a$, because $u=v=0$. The expression for 
the Hamiltonian is 
\beq
H=\beta \sqrt{m^2c^4+(c\overrightarrow{p} - e\overrightarrow{A})^2
- \hbar ce\,\overrightarrow{\Sigma}.\overrightarrow{B}}\,.
\eeq
This is exactly the result obtained by Eriksen and Kolsrud
\cite{EK} for this case.

%%%%%%%%%%%%%%%%%%%%%%%%%%%%%%%%%%%%%%%%%%%%%%%%%%%%%%%%%%%
%%%%%%%%%%%%%%%%%%%%%%%%%%%%%%%%%%%%%%%%%%%%%%%%%%%%%%%%%%%
\section{Particle with anomalous magnetic moment in a
static magnetic field}

In this case we start from the Hamiltonian with 
$$
K^a_b=-i\,\hbar \,c \,\delta^a_b
$$ 
and $g_a = -eA_a +\alpha_a \mu_I\overrightarrow{\Sigma}.
\overrightarrow{B}\,,$ and again with $u=v=0$. 

This version is a bit more complicated because one has to 
account for the commutators of $g_a$ with $\al^b$ and 
$\beta$. After some calculations we arrive at the 
following form of the Hamitonian 
\beq
H^2 = m^2c^4 + (c\overrightarrow{p} 
- e\overrightarrow{A})^2
- 2 \mu_{I} mc^2 \overrightarrow{\Sigma}.\overrightarrow{B}
+\mu_I^2B^2 + \mu_I \beta 
\overrightarrow{\Sigma}
\,\cdot\,
\left(\overrightarrow{B}\times\overrightarrow{p}-
\overrightarrow{p}\times\overrightarrow{B}\right)\,.
\eeq
Once again, this expression is in a perfect agreement with the 
result obtained by Eriksen and Kolsrud \cite{EK}.

%%%%%%%%%%%%%%%%%%%%%%%%%%%%%%%%%%%%%%%%%%%%%%%%%%%%%%%%%%%%%%


\begin{thebibliography}{99}
\bibitem{notations}
Our notations are $i,j,k, ...=0,1,2,3\,$ and $a,b,c,...=1,2,3$; 
$\eta_{ij}\,=\,\mbox{diag} (-1,\,1,\,1,\,1)$; the 3-dimensional indices 
$a,b,\dots$ are raised and lowered with the help of the Euclidean metric 
$\delta_{ab}$.
\end{thebibliography}
\end{document}